# Particle Tracking, Recognition and LET Evaluation of Out-of-Field Proton Therapy Delivered to a Phantom with Implants

Cristina Bălan[1,2], Carlos Granja[3], Gennady Mytsin[4], Sergey Shvidky[4], Alexander Molokanov[4], Lukas Marek[3], Vasile Chiş[1], Cristina Oancea[3*]

1. Faculty of Physics, Babeș-Bolyai University, Cluj-Napoca, Romania
2. Department of Radiotherapy, The Oncology Institute "Prof. Dr. Ion Chiricuta", Cluj-Napoca, Romania
3. ADVACAM, Prague, Czech Republic
4. International Intergovernmental Organization Joint Institute for Nuclear Research (JINR), Dubna

**\*** Corresponding author: cristina.oancea@advacam.cz

**ABSTRACT**

**Objective:** This study aims to assess the composition of scattered particles generated in proton therapy for tumors situated proximal to titanium dental implants. The investigation involves decomposing the mixed field and recording Linear Energy Transfer (LET) spectra to quantify the influence of metallic dental inserts located behind the tumor.

**Approach:** A therapeutic conformal proton beam was used to deliver the treatment plan to an anthropomorphic head phantom with two types of implants inserted in the target volume (made of titanium and plastic, respectively). The scattered radiation resulted during the irradiation was detected by a hybrid semiconductor pixel detector MiniPIX Timepix3 that was placed distal to the Spread-out Bragg peak. Visualization and field decomposition of stray radiation were generated using algorithms trained in particle recognition based on artificial intelligence convolution neural networks (AI CNN). Spectral sensitive aspects of the scattered radiation were collected using two angular positions of the detector relative to the beam direction: 0° and 60°.

**Results:** Using AI CNN, 3 classes of particles were identified: protons, electrons & photons, and ions & fast neutrons. Placing a Titanium implant in the beam's path resulted in predominantly electrons and photons, contributing 52.2%, whereas for plastic implants, the contribution was 65.4%. Scattered protons comprised 45.5% and 31.9% with and without metal inserts, respectively. The LET spectra was derived for each group of particles identified, with values ranging from 0.01 to 7.5 keV·$\mu m^{-1}$ for Titanium implants/plastic implants. The low-LET component was primarily composed of electrons and photons, while the high-LET component corresponded to protons and ions.

**Significance:** This method, complemented by directional maps, holds potential for evaluating and validating treatment plans involving stray radiation near organs at risk, offering precise discrimination of the mixt field, enhancing in this way the LET calculation.

# 1. Introduction

Proton therapy is strongly indicated, over the photon treatments, for head and neck cancer (HNC) patients for two major reasons: physical properties of dose deposition and the Bragg peak. This makes possible a unique way of dose collimation to the tumour shape for a drastic dose reduction to the critical organs at risk that surrounds the tumour site, sparing the healthy tissue and reducing the toxicity level (Blanchard et al., 2018; Leeman et al., 2017; Ramaekers et al., 2013). For patients with dental implants, the treatment should be handled differently compared with the standard cases. The medical decision assumes radiotherapy treatments for patients with various types of high-density materials in their bodies, representing a challenging balance between benefits and clinical outcomes (Müller et al., 2020a). Most dental inserts used nowadays have titanium (Ti) in their components, or they are Ti-based alloys (Datte et al., 2018; Kim et al., 2019). The scattered radiation produced by the implants can induce perturbations into irradiated volume, planning dose calculation and delivery (Reft et al., 2003; Rousselle et al., 2020), creating complications in achieving clinical outcomes (Jia et al., 2015). Approx. 4% of the oncological patients following radiotherapy treatments present high Z-regions in the irradiated volume (Le Fèvre et al., 2022; Reft et al., 2003). Despite the material development in medical science, as the discover of the carbon fiber or peek dental implants (polyetheretherketone) used in dental reconstruction (Akyol et al., 2021, 2019; Lommen et al., n.d.), there are situation when the metallic inserts cannot be removed, being centered in the target volume. For this kind of cases, the direct avoidance of those non-tissue equivalent materials is difficult (Acquah et al., 2018; Jäkel and Reiss, 2007; Müller et al., 2020b).

During proton irradiation, those inhomogeneities are expected to create a significant impact on the irradiated volume compared to the well-studied effects reported in photon beams (Azizi et al., 2018) (Çatli and Tanir, 2013). With a considerable artefact at the metal/tissue interface, there is no evidence of a proper dose calculation: up to 10% difference has been reported in a dosimetric study in the treatment planning process for patients with chordoma that possess Ti implants even at 10 mm away from the insert's location (Verburg and Seco, 2013). Metallic screws are frequently used in spinal pathologies, so measuring a dose near such implants in a proton-based treatment, an overestimation of 12%, respective 8% for stainless steel and Ti screws, were reported. They also described discrepancies between dose calculations provided by the TPS and those that were measured: the calculation showed a 9% enhancement in dose within 1 cm from the lateral interface of the insert, but the measurement concludes an increased dose only for 5 mm away from the screws (Jia et al., 2015). Concerning the effect of secondary particles on linear energy transfer (LET) values behind hip implants, the impact of secondary particles with and without metallic implants has been investigated, and a multiplication of 9, respectively 6 times the absorbed dose after Ti and stainless-steel hip prostheses was reported when implants were closed to the Bragg peak (Oancea et al., 2018a).

A previous experiment that investigated the effect of metallic dental implants by evaluating the 2D distribution of the absorbed dose near the implants motivated us to move further and thoroughly characterize the particle field behind the tumour and identify the type of scattered radiation, including scattered protons, photons, electrons, and ions (Oancea et al., 2017). Two limitations were highlighted: particles that have a LET less or around 7 keV/μm and electrons cannot be detected by the dosimetry system used in that experiment (Oancea et al., 2018a, 2018b, 2017). Presented as a new generation of pixeled detectors with a novel design, using an active semiconductor, MiniPIX Timepix3 detector with an ASIC chip, allows particle discrimination and real time visualization of those mixed fields of stray radiation, detecting particles with LET values starting form 0.1 keV/μm to over 100 keV/μm (Charyyev et al., 2020; Granja et al., 2021, 2018a; Nabha et al., 2022; Novak et al., 2023; Oancea et al., 2023a). Besides Timepix3 detectors, there are other types of detectors (passive or active) that can be used for LET measurements, for example, solid-state nuclear track etched detectors (Oancea et al., 2017), Liulin detectors, tissue-equivalent proportional counters (Magrin et al., 2023) and optically stimulated luminescence (Granville et al., 2016). However, they cannot detect all types of charged particles, not being able to create an appropriate decomposition of the stray radiation. Because of this, the measured LET spectra with these detectors is maintained in a limited range, even more the data evaluation becomes time-consuming.

Using protons or other types of heavy ions as a therapeutic modality, the radiobiological effectiveness should be considered in the final dose calculation. With an increased ionization density produced by the charged particles in irradiated medium, the linear energy transfer concept was generally accepted by the scientific community to describe the biological effects on the healthy tissue (Grassberger and Paganetti, 2011). Nowadays, modern treatment planning systems used in proton or heavy ion facilities include in the final dose optimization process a LET based model (An et al., 2017; Gu et al., 2021; Liu et al., 2013). The LET of the particles that are crossing a pixeled detector is of interest for primary and secondary radiation spectral characterization especially when recent developments of TPS allow for LET treatment planning (McIntyre et al., 2023; Nabha et al., 2022; Novak et al., 2023; Stasica et al., 2023).

The goal of this study is to characterize the scattered particles produced by a proton beam using a head and neck case, with two metallic inserts placed in the beam's path. The investigation is focused on a single field treatment plan delivered to an anthropomorphic head phantom, wherein the target volume incorporates two Ti implants situated in center of the target volume. An identical procedure was conducted to establish a thorough analysis, replacing the Ti implants with plastic substitutes.

In this work, an experimental setup using a MiniPIX Timepix3 detector strategically placed beyond the Spread-out-Bragg-Peak (SOBP) was designed to evaluate the type of scattered particles and their contribution to the LET spectra. This placement facilitates an in-depth exploration of the effects of scattered radiation and the subsequent generation of secondary particles, including neutrons. We performed single particle tracking and provided spectral information at a microscopic scale for individual particles, quantifying the impact of secondary particles behind Ti implants. This approach comprehensively quantifies the influence exerted by secondary particles in the vicinity of Ti implants.

## 2. Method

### 2.1. Proton beamline

Measurements were carried out using a therapeutical proton beam of 170 MeV energy, delivered using a double scattering system. The almost monoenergetic beam with a wide beam size of 8x8 cm$^2$ homogeneous in cross-section was shaped for the specific treated case using corresponding collimators and boluses manufactured according to the standard process for a patient. The Bragg peak was spread out with a ridge filter, creating the so-called "Spread Out Bragg Peak" (SOBP) of about 2 cm length.

### 2.2. Anthropomorphic head phantom

This work aims to reproduce a realistic head and neck treatment using an anthropomorphic head phantom with human-like anatomical densities. Irradiating a phantom with complex heterogeneities in its construction creates a field with multiple sources of scattered radiation. The Alderson phantom (Radiology Support Devices, Inc., CA, USA), see figure 1 (a), was used in this study.

The experimental setup involves two distinct scenarios: one with the phantom containing titanium metallic implants and the other utilizing tissue-equivalent (TE) inserts. The metallic inserts are dental implants used for tooth replacement. Two cylinders made from titanium (density of 4,51 g/cm$^3$), with a length of 1 cm and a diameter of 0.33 cm, were inserted in the phantoms' upper jaw, to reproduce the replacement of molars. Figure 1 (b) shows the metallic implants used in this experiment and their position inside the head phantom.

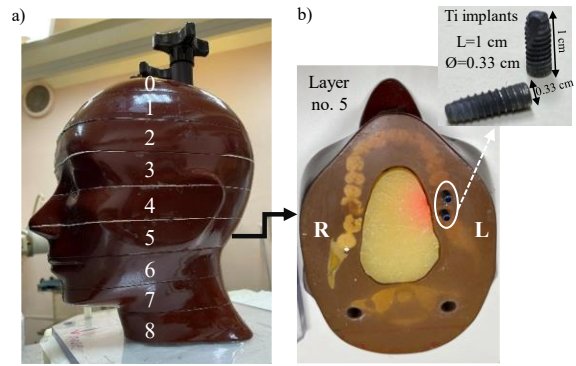

*Figure 1 a)* *The anthropomorphic head phantom (Alderson phantom) made out from 9 independent layers that sum up the whole anatomy of the human head;* ***b)*** *two cylindrical metallic implants made from titanium (Ti) were inserted into the 5$^{th}$ layer of the Alderson phantom, at the molars' level.*

### 2.3. Treatment planning

The anthropomorphic phantom was scanned using an X-ray CT scanner, and the obtained CT images were used to create a 3D treatment plan. The target was contoured, so the planning target volume (PTV) contained the Ti implants at its center. A pencil-beam algorithm and an in-house design treatment planning system (TPS) were used to create a 3D plan, using only one conformal proton field. The SOBP was planned to be inside the PTV, including the implants. A collimated proton beam of 170 MeV, directed perpendicular to the anthropomorphic phantom, was delivered to the phantom (irradiation angle: 90°). For this experiment, the treatment plan contoured on the X-ray CT was used only for conformational reasons, and the planned dose and its delivery were not prioritized.

## 2.4. MiniPIX Timepix3 detector

Hybrid semiconductor pixel detectors have been used in this study to evaluate the impact of metallic implants placed into an anthropomorphic phantom in a proton beam treatment. The MiniPIX Timpex3 (TPX3) detectors (ADVACAM s.r.o), with the Timepix3 chip developed by the Medipix Collaboration (from CERN) were used. The readout system is formed by an ASIC chip with a square matrix of 256 x 256 pixels, having a total of 65536 independent channels responsible for collecting data. Each pixel contains its own electronics, and it has a pixel size of 55 x 55 µm$^2$, resulting in an active area of 14.08 mm$^2$. Timepix3 detectors can collect synchronized information about energy, time, or event numbers (Llopart et al., 2007; Poikela et al., 2014) having two channels dedicated to each pixel.

## 2.5. Detectors operation and data processing

Pixet Pro v.1.8.0., provided by ADVACAM, is the dedicated software package used in detector operations, data management and online output visualization (ADVACAM.). A 2.0 USB cable provides the connection between the detector and the PC where the software is located.

Simultaneous measurement of energy and time interaction is possible due to the design of the detector, meaning that the TPX3 readouts collect information regarding Time over Threshold (ToT)/energy mode and Time of Arrival (ToA)/time of individual particles. Applying calibration for each detector's pixels, the ToT data is converted into deposited energy (Jakubek, 2011). The calibration process also involves equalization, where threshold for each pixel was adjusted, offering enhanced detection efficiency for charged particles. For the ToA signal registration, a digital counter is dedicated to each pixel, in addition to the components responsible for ToT signal collection (preamplifier and discriminator), that starts to count the time elapsed from the first hit in the pixel in a specified time window, with 1.56 ns resolution (Granja et al., 2018a; Oancea et al., 2023a; Turecek et al., 2016). When a charged particle interacts with the pixelated sensor of the TPX3, its energy diffuses in both directions, lateral and horizontal, producing a particular track for that particle called "cluster". Having this kind of infrastructure, the semiconductor Timepix-based detectors can discriminate different particle types based on the properties and morphology of each cluster (Vykydal et al., 2006, Oancea et al 2023).

Data processing was done using the Data Processing Engine (DPE) tool (Marek et al., 2023). Using integrated Phyton scripts, this software can process data measured with Timepix3 detectors giving information about clusters, their morphology, directional or spectrometric parameters collected by the pixeled matrix. Based on the artificial intelligence, a trained neuronal network (NN) could discriminate the particle types in a mixed radiation field (Marek et al., 2023, Granja et al., 2018; Oancea et al., 2023).

### 2.6. Experimental setup

The anthropomorphic phantom was placed directly into a collimated proton beam of 170 MeV. To analyze the impact of the metallic implant on a proton-based treatment plan, two irradiations were performed: one with metallic implants in PTV and the other with TE replacement implants. Based on the plan that was calculated in the TPS, the SOBP was in the PTV area. The primary protons exiting the beam nozzle interact with customized boluses and collimators to shape the beam according to the tumor dimensions.

An inline setup was chosen, placing the active region of the detectors behind the head phantom as seen in figure 2, beyond the PTV. The scattered radiation produced by the interaction of the primary protons with the phantom in each situation was registered and monitored by one detector in miniaturized architecture (MiniPIX) with a Timepix3 chip and a silicon (Si) sensor of 500 μm. The Bragg peak was planned so that its position should be in the interest region, where the metal inserts are located. The detector was placed 9 cm distal to the Bragg Peak, near the phantom. Reducing the incident proton beam intensity allowed for single particle tracking response and identification resolving power in both situations. Thus, the composition of the radiation field was analyzed event by event. Based on their characteristic directional-sensitive spectral-track signal in the detector, particles like protons, photons, electrons, and ions were identified in the stray radiation (Marek et al., 2023).

A schematic representation with a photo of the experimental setup in the treatment room is displayed in figure 2.

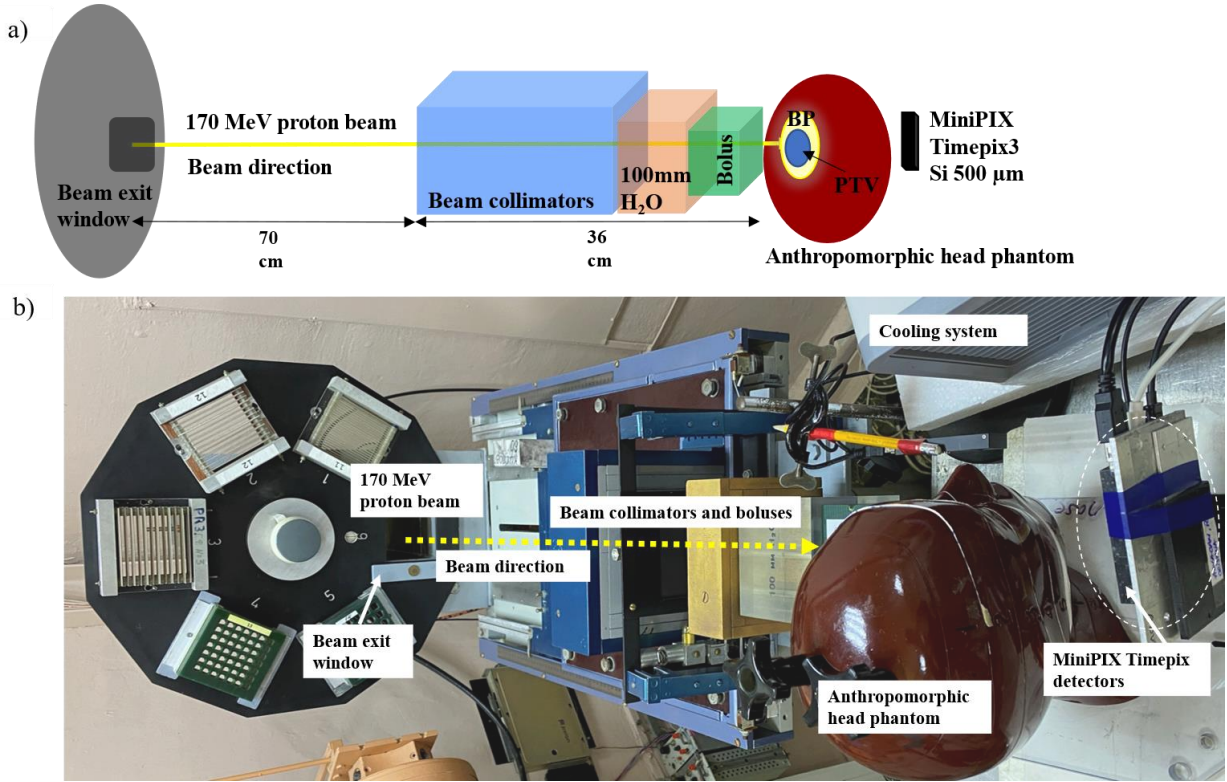

*Figure 2. Experimental setup at the therapeutic proton beam line. **a)** schematic illustration of the experiment: a fixed proton beam with a nominal energy of 170 MeV is entering the treatment room. The beam is planned to deliver the treatment based on a treatment plan to an anthropomorphic phantom. The conformity is achieved in the PTV based on the beam collimators, one block of virtual water equivalent to a 100 mm $H_2O$ plus, a personalized bolus and collimator. **b)** a photo from the treatment room with the setup: an incident proton beam with a nominal energy of 170 MeV was used to deliver the treatment to the phantom. Detectors arrangements were done in such manner as the scattered radiation produced beyond the SOBP could be measured.*

### 2.7. LET conversion from Si to water

The equation used in LET in silicon (LET(Si)) calculation represents the ratio between deposited energy of the particle, dE, and the path of the cluster through the Si sensor in 3D, dx = $L_{3D}$. The result represents the magnitude of the LET specific for each event that occurred in pixeled matrix of the Si sensor (Granja et al., 2021). The LET (Si) is determined based on the following equation:

$$LET = dE/dx \; [\text{keV/µm}] \quad (1)$$

Tracking aspects of the cluster is a combination of two directional paths: one alongside of the longitudinal trajectory on the sensor surface, a 2D length, and another one through the transversal section of the Si sensor. The 3D length (L$_{3D}$) of the tracked event is defined based on the Pythagoras formalism:

$$L_{3D} = \sqrt{Length_{2D}^2 + (sensor\ thinckness)^2} \quad (2), where$$

$$Length_{2D} = L_{2D} - 2.5 \cdot \sigma_a \quad (3)$$

The 2D trajectory projection with its standard deviation, $\sigma_a$, along with the known sensor thickness, determines the directional contribution for the LET (Si) calculation (Granja et al., 2021; Oancea et al., 2023a).

Following the recommendations stated in ICRU 78, TG 224 and from the code of practice (IAEA-TRS 398) often used in clinical practice in radiotherapy, the calculation of the absorbed dose should be determined in water (Arjomandy et al., 2019; IAEA, 2000; ICRU 59, 1998). During this experiment, the scattered radiation was characterized using a pixeled detector with a Si sensor. Thus, the measured LET(Si) was converted to LET in water (LET(water)) using the fitting function proposed by *Benton et al.* (Benton et al., 2010).

## 3. Results and Discussion

### 3.1. Mixt field characterization and particle discrimination based on CNN model

Figure 3 represents a 2D map of the per pixel deposited energy at the sensors' level produced by 2000 particles from the scatter radiation after the interaction of the collimated proton beam with the targets. Both situations, with Ti implant (first row) and TE inserts (second row) are presented using two angular positions of the detector related to the proton beam axis. The detector was placed perpendicular to the direction of the incident proton beam , referred as 0° in the next paragraphs, respectively 60° relative to the beam direction. Using an angular position for the sensor plane, the spectral resolution of the sensor is increased so the event discrimination and characterization is significant improved (Granja et al., 2018a).

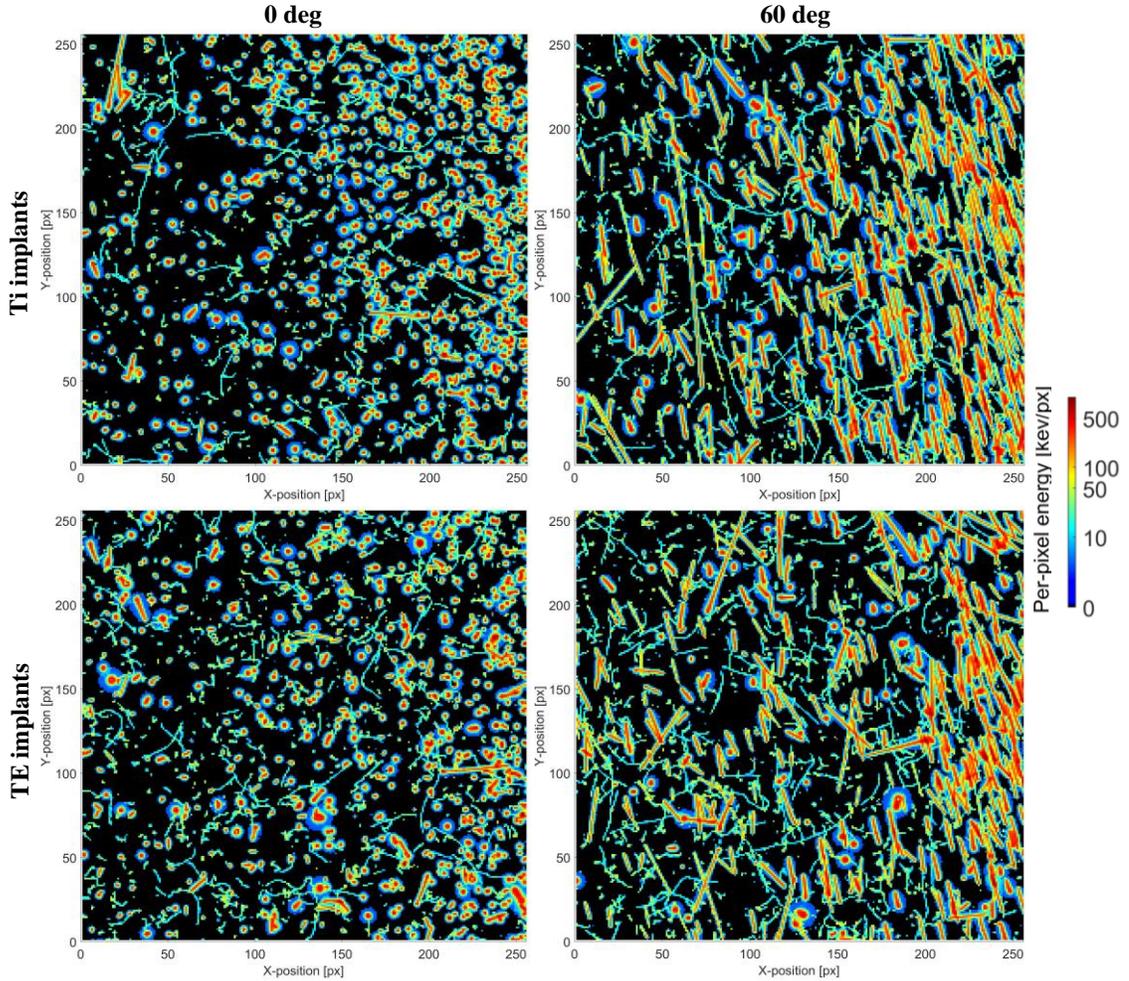

*Figure 3. Deposited energy of selected first 2000 particles from the scattered radiation produced by the proton in both scenarios: with Ti implants (**first row**), and with plastic inserts (**second row**) are presented as 2D maps of the whole sensor, in a color logarithmic scale. Mixed field of particles created after the primary proton beam interactions with the phantom created in 14.08 x 14.08 cm$^2$ area of the MiniPIX Timepix3 with a Si sensor of 500 µm, placed at two different angles: 0° (**left**), respectively 60° (**right**). Particles tracks of incident particles create characteristic clusters on the sensor surface, according to their energy and type.*

Despite the presence of the implants, the incidents particles from the stray radiation create distinctive clusters on the active surface of the TPX3 detector. The spectral tracking of the particles is based on two general concepts: the color scale of the deposited energy per pixel, and the trajectory morphology respectively. Tracks with deposited energy of 500 keV/px and above are particular to the high energetic particles (protons, ions, fast neutrons), while the tracks with lower-deposited energy (e.g. between 3-200 keV/px) represent the specific track for light particles (photons, electrons). Considering the morphology of the track, the irradiation geometry, and the detector angular position relative to the incident beam, the mixed radiation field is very complex.

Using trained artificial intelligence (AI) algorithms based convolutional neural networks (CNN) models from DPE (Marek et al., 2023), the decomposition of the mixed particle field from the scattered radiation was done and is presented in figure 4, in both situations: with Ti (right) and TE (left) implants. The first 200 events from the mixed field are shown in figure 4a. Using CNN (Marek et al., 2023) those 200 particles in each case are deconvoluted in three classes based on the pattern reorganization cluster and their directional, spectroscopic, and morphological properties: i) protons (figure 4b), ii) electrons and photons (figure 4c), iii) ions and fast neutrons (figure 4d).

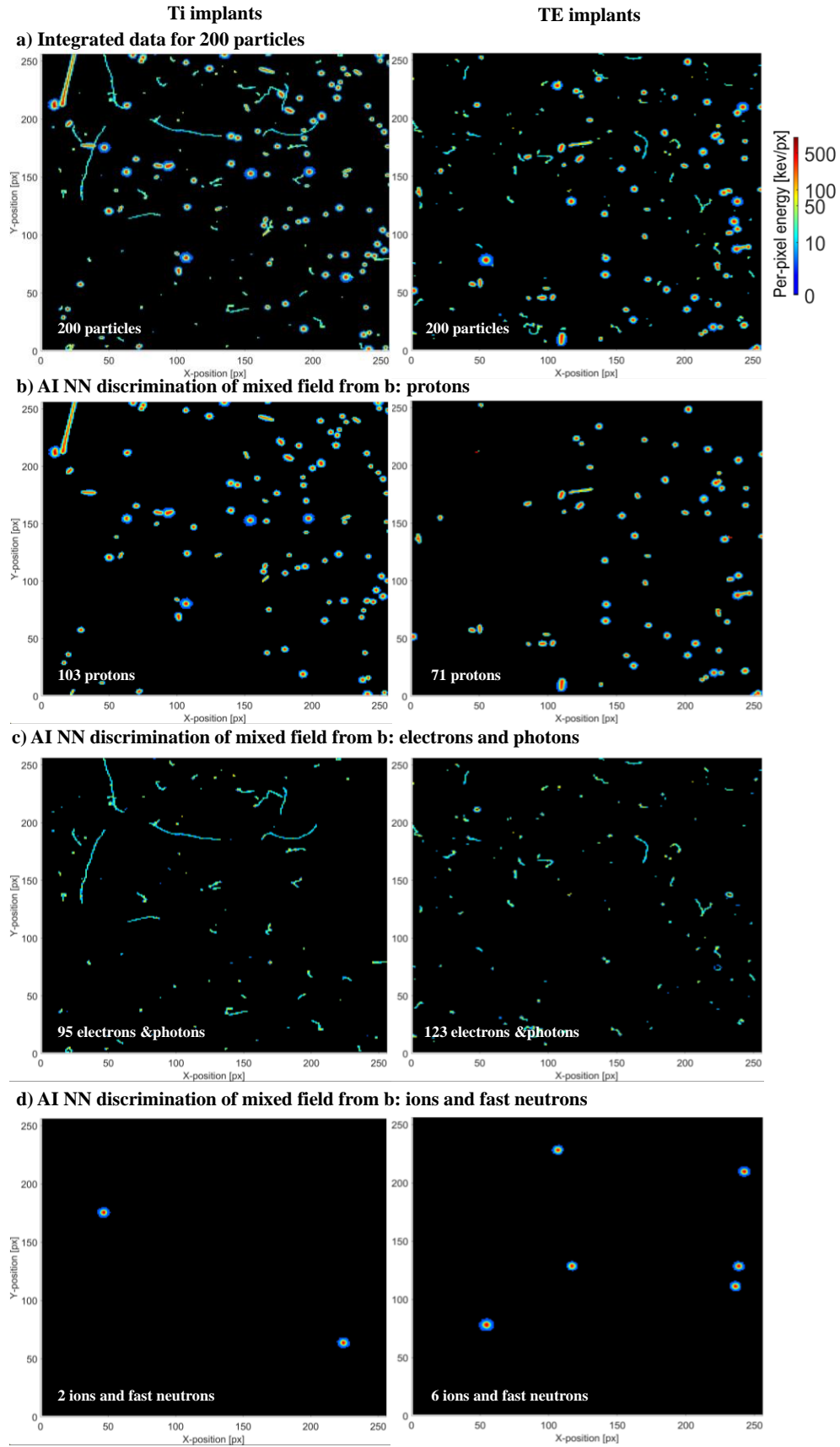

*Figure 4. The spectroscopic detection of the scatter radiation with Ti implants (**first column**) and with tissue equivalent (TE) inserts (**second column**) produced by a 170 MeV proton beam at the full pixel matrix (256 x 256 pixels) of the MiniPIX TimePix3 detector with a 500 µm Si sensor thickness inclined at 0°. a) Displaying 200 particles from all integrated data measured by the Si sensor in both scenarios presented in figure 3, with a perpendicular position of the sensor (0°). Using particle identification algorithms, a trained AI neuronal network from DPE tool, the scattered radiation shown in a) is discriminated based on particle's type as following: protons (b), electrons and photons (c), ions and fast neutrons (d).*

The percentage composition of scattered radiation resulted from the anthropomorphic phantom for each particle groups previously listed, for two positions of the sensor relative to the beam axis: 0° and 60° (from figure 3 and 4, all measured particles) were quantified in table 1. Measuring the scattered particle field caused by the interaction of the proton beam with the phantom in both angular distributions showed specific composition of the scatter radiation as a function of the pixel matrix orientation. In table 1, all data were used to provide the percentage composition of the mixed field. Scattered protons were more prominent with Ti implants compared to TE implants in both measurements, with approximately 13% greater detection for the Timepix3 perpendicular to the beam direction and 8% more for the other angular position. On the other hand, more electrons and photons were detected behind the TE implants. The contribution of ions and fast neutrons at the sensor level is very low, below 1% at the 60° and above 2% for 0°. Notably, no thermal neutrons were measured at the detector position. Based on the limited Si sensors response to the neutron's detection some dedicated converter should be placed on the active surface (Granja et al., 2023b; Oancea et al., 2023b). For this reason, this study presents some limitation regarding the detection and spectral characterization of neutrons where only direct interactions in Si were detected.

*Table 1* *The particle type decomposition of the scattered radiation produced by the incident proton beam measured with the TPX3 detector with 500 µm Si sensor with Ti and TE implants inserted in the anthropomorphic phantom is shown in percentages (%).*

| | Sensor orientation | 0° | | 60° | |
|---|---|---|---|---|---|
| | Implant material | Ti | TE | Ti | TE |
| Particle groups | protons | 45.5 ± 1.0 % | 31.9 ± 0.3 % | 28.0 ± 0.6 % | 20.3 ± 1.0 % |
| | electrons and photons | 52.2 ± 0.9 % | 65.4 ± 0.1 % | 71.2 ± 0.5 % | 78.8 ± 0.9 % |
| | ions and fast neutrons | 2.3 ± 0.2 % | 2.7 ± 0.3 % | 0.8 ± 0.1 % | 0.9 ± 0.1 % |

### 3.2 LET spectra of scattered radiation

Measuring the LET spectra of the scattered radiation produced by a collimated proton beam, when the two Ti implants are placed in the isocenter compared with a measurement that have two plastic inserts, made from tissue equivalent material, highlight the impact of those inserts in this proposed setup. To access the biological signification of these measurements distal to the SBOP, the LET in water was derived to emphasize the impact on the healthy tissue. Therefore, the LET in water values for both setups are presented in figure 5. The LET spectra were recorded for two angular positions of the TPX3 detector: 0° (red) and 60° (blue) to cover a wide field-of-view and to increase the discrimination power of the detector (Granja et al., 2021, 2018a). Data was normalized to the maximum value for all particles, for both setups. The LET spectra in figure 5a shows the contribution of all particles with Ti implants (solid line) and TE inserts (translucent line). The contribution of individual particle classes to the total LET spectra is represented separately as follows: protons (figure 5b), electrons and photons (figure 5c), and ions (figure 5d). In agreement with the percentual composition presented in table 1, secondary and scattered protons are bringing the largest contribution to the LET spectra with values ranging from 0.5 to 7.5 keV·µm$^{-1}$ and a maximum value at 1.35 keV·µm$^{-1}$ for both setups, with Ti implants and with plastic inserts at 0° configuration. Placing the detector at 60° relative to the incident proton beam direction increases the particle path across the sensor thickness, resulting in a wider deposited energy distribution (Nabha et al., 2022). As figure 5b shows, the maximum value of derived LET(water) for protons that are crossing the sensor placed in a larger incident angle is shifted to 2 keV·μm$^{-1}$. Directional detection, together with angular resolution of this setup a second peak at 4.5 keV·μm$^{-1}$ can be seen observed at the end of the LET(water) spectra, weighting the non-negligible importance of the particle length into the cross section of the sensor. The proposed methodology gives some advantages regarding LET(water) characterization by using large angular position for the detectors to obtain the best tracking and spectral analysis to calculate the LET values. Out-of-field radiation measured after the SOBP, produce multiple secondary protons in a considerable range of LET(water) values. Labeled generally as 'protons' by the CNN infrastructure, this group of single protons are described by a wide range of LET(water) values from 0.5 keV·μm$^{-1}$ up to 7.5 keV·μm$^{-1}$. The main goal of this study is to highlight the discrepancies between measurements performed with metallic inserts placed in the beam's path, respectively without them. The LET(water) spectra for protons exhibit the same shape and interval values in both scenarios, regardless of the presents of the metallic inserts into the target volume. There was no shift detected in the maximum values, meaning that the LET distribution was not impacted by the metal inserts. Without any spectral discrepancies of the LET distribution, the present of the Ti implants produces an amplification of statistical events up to 30% of single protons detected with a LET(water) value closed by its maximum were measured by the Timepix3. Further investigation should consider the impact of high energy protons and their effects on the healthy tissue, including an estimation of their impact on the total dose deposited beyond the PTV.

The contribution of electrons and photons to the LET(water) spectra is shown in figure 5c and it exhibits identical shape and values for both setups with and without implants. Known as light particles, with a very short stopping power range, these particles are characterized by a low LET(water) (Vassiliev et al., 2018). With LET(water) values below 1 keV·μm$^{-1}$ for 60° position of the detector, 1.5 keV·μm$^{-1}$ respectively for perpendicular position relative to the beam direction, no differences were seen when comparing the setups with Ti implants and without them. Even though special attention is needed to the specific particle's interaction between incident proton beam and this high Z-materials. A detailed analysis of these interactions is presented in the following paragraph.

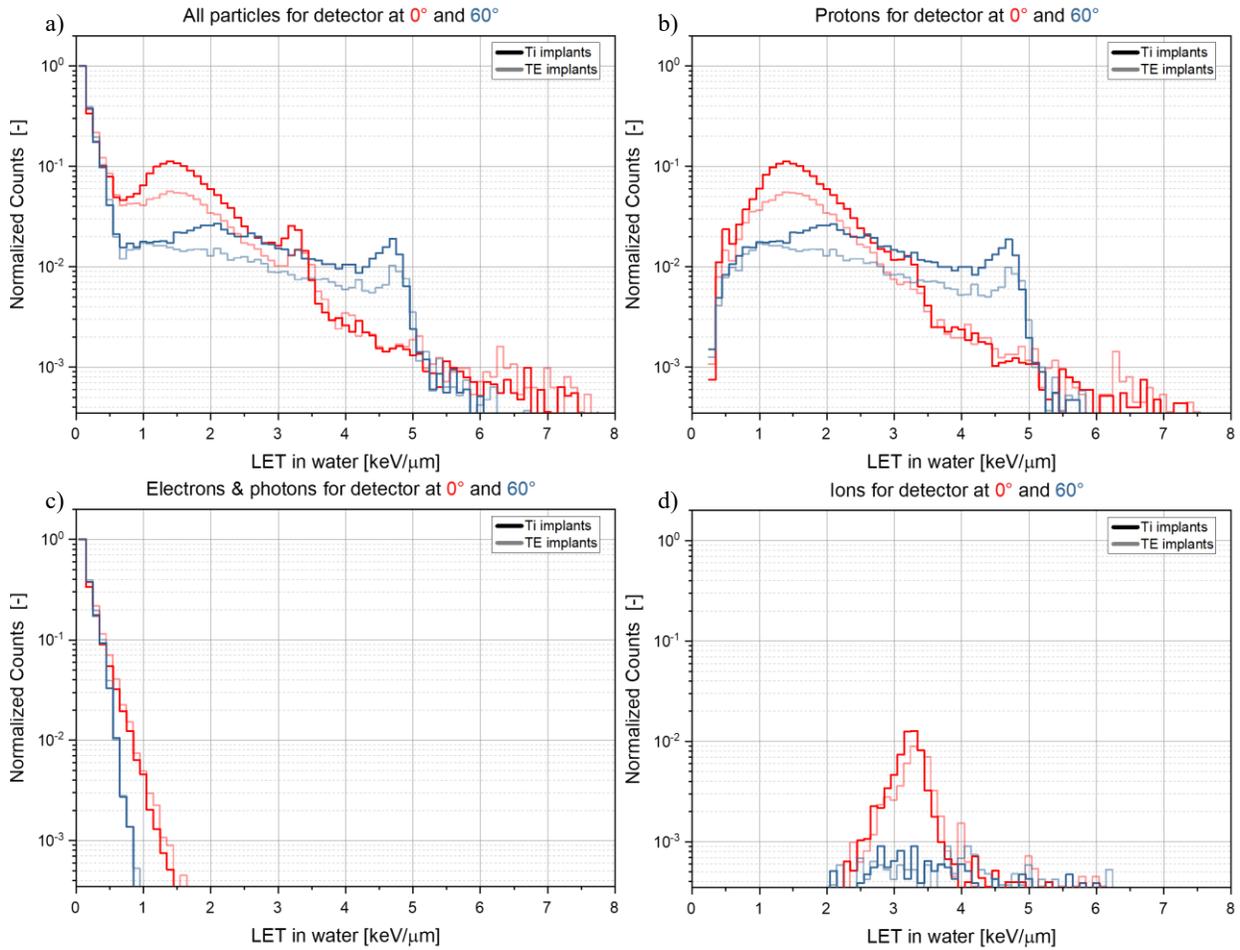

*Figure 5. LET spectra (LET water) of the scattered radiation produced by an incident 170 MeV proton beam passing the anthropomorphic phantom with two different materials for dentals implants: continuous line used for Ti implants results, and dash line for the tissue equivalent (TE) plastic implants in two angular position of the TPX3 detector: 0° (red), respectively 60° (blue); a) LET spectra for 35k particles from mixed scattered radiation; LET spectra for the results of AI NN particle discrimination algorithm:(b) protons, (c) electrons and photons , (d) and ions. Data was collected using one TPX3 detector, with 500 μm Si sensor.*

Following the mechanism of the proton's interaction with matter, the energetic protons used in radiotherapy at the Bragg peak are implicated in nuclear interactions and inelastic collision (Matsuzaki et al., n.d.; Thompson et al., 2015). The products resulted from this interaction of the primary protons with anatomically densities from the anthropomorphic head phantom when no metallic inserts are placed in the beam's path are represented by the secondary electrons and prompt photons (Matsuzaki et al., n.d.; Zarifi et al., 2017). With the same LET value, the ratio between the primary protons and secondary negative charged particles, together with photons is not following the same trend. A Monte Carlo simulation of a proton beam that is passing through a titanium implant highlighted that only a small fraction of the incident protons interacts *via* nuclear interaction with the implant, and the multiple Coulomb scattering of the protons that are passing the medium are responsible for most uncertainties that are reported in proton therapy when Ti implants are inserted in the irradiated medium (Verburg and Seco, 2013). These computational results that were reported, agree the experimental results obtained in this experiment, where the total count of protons from the out-of-field radiation is statistically increased, compared to the constant number of events for electrons and photons when the Ti inserts are present.

Ions are the third class of particles that CNN algorithms were identify in the stray radiation field, and they are the nuclear products of the nuclear interaction between source protons with the target volume in the Bragg peak region (Thompson et al., 2015). The contribution of ions to LET spectra is presented in figure 5d where the value of the derived LET in water ranges from 2.0 keV·µm$^{-1}$ to 6.3 keV·µm$^{-1}$. Overall, the presence of the Ti implants does not lead to any enhancement of the ion's contribution, at the position where data were collected and presented in this study.

Furthermore, the work is focused on the impact of particles from the scattered radiation on the healthy tissue, that are characterized by a high value of the LET. Considerable attention should be paid to an extended analysis of protons by creating a classification based on the spectral aspects of each proton. Using directional information, in addition to the spectral and morphological analyzes proposed in this study and presented in the first part of this work, we also extend the methodology by adding directional maps of two classes of protons: high and low energy protons.

**3.3. Scattered proton: spectral-sensitive tracking and directional maps**

Exploiting the per-pixel spectral sensitivity and tracking response of the pixel detector (Granja et al., 2018b), directional maps of scattered protons were created. Two groups of protons were identified in the mixed radiation and selected to be used to plot their directional distribution of as seen in figure 6, namely class1: low-energy protons and class2: high-energy protons. The events in the data shown correspond primarily to tracks produced in the detector by scattered protons. To a lesser extent, part of these events in the measured data in this work, namely short-range high energy loss tracks, are produced also by fast neutron induced interactions in the detector silicon sensor (Granja et al., 2023a). A more quantitative discrimination of events between these two components, protons, and neutrons, requires further and detailed analysis including detector specific response matrices derived from dedicated calibrations in well-defined neutron fields (Granja et al., 2023b).

The direction of incidence can be derived and reconstructed from the detailed tracking registration (as seen in figure 3). Resulting maps of angular flux (Granja et al., 2018b) of distinct particle-directional groups were produced and shown in figure 6. The data shown was collected in longer time intervals: 16.2 min and 18.1 min for the measurements with the implants and without the implants. The events displayed were filtered from a total of 5769 and 2621 particles, respectively. The primary beam intensity was partly different resulting in different particle rate on the detector: 135 and 71 particles·$s^{-1}$·$cm^{-2}$. The results can be normalized to each other by the different beam intensity by a factor 71/135.

The resulting distributions and patterns observed in the spatial and directional (figure 6) maps are determined by convolution primarily of the beam morphology (size, intensity), the phantom morphology (size, shape, material, density). In part also by the geometry and space in between the accelerator beam nozzle and the phantom, and between the phantom and the detector. Patterns and specific differences in the distributions can be observed such as the lateral spread and the relative yield along the horizontal and vertical directions can be partly correlated to the presence or absence of the implant.

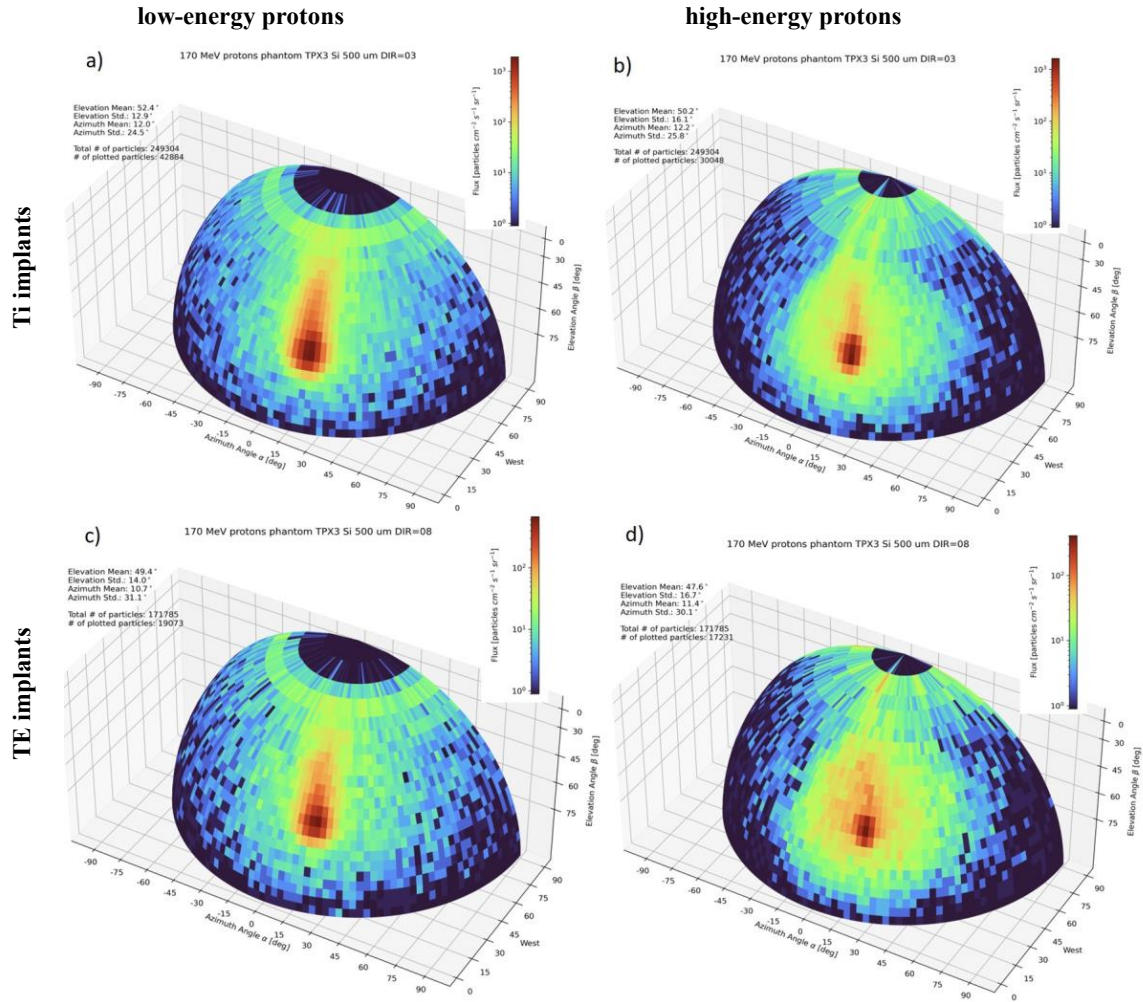

*Figure 6. Directional maps of incident scattered protons selected from the data given in Fig. 3, for an angular position of 60° relative to the incident beam line. Results are given for data with the head phantom equipped with Ti implant (top row) and without (bottom row) the Ti implant (TE implants). The two selected spectral-tracking groups are shown: low-energy scattered protons (left column) and high-energy scattered protons (right column).*

## 4. Conclusions

In this work, the impact of dental implants made of Ti alloy in a proton-based treatment using an anthropomorphic head phantom is quantified. Timepix3 detector with a Si sensor was used to measure and characterize a mixt radiation field produced by the interaction of a conformal proton beam with a nominal energy of 170 MeV with Ti implants placed in the lower mandibular region of the head phantom. Repeating the measurements with the same setup, but without metallic inserts (replaced by tissue equivalent implants) offered a comparative perspective of the dosimetric parameters in both scenarios. Comparable situation with a real head and neck cancer treatment, this study proposed a detailed analysis of the impact of the metallic inserts into the proton beam by evaluating the LET distributions and their compositions in terms of particle type. The particle identification and discrimination were done using trained neural networks for individual track recognition.

The scattered radiation decomposition in both scenarios revealed a contrast in proton and light particle as electrons and photons ratios. Using two angular positions for the sensor orientation relative to the incident proton beam direction, the mixt field was classified based on their specific track in three classes: protons, electrons with photons and ions. Placing Ti implants into the incident beam, the ratio between scattered protons (45.5%) and electrons with photons (52.2%) is almost double (ratio Ti/TE= 0.87) then that one obtained when the TE implants were replacing the metallic ones, 31.9% scattered protons with 65.4% light particles (ratio Ti/TE = 0.48 ). The higher detection of scattered protons by the TPX3 detector when Ti implants are used, resulted from the metal's interaction with the incident proton beam.

Using two angular positions for detector (0° and 60°), the high discrepancies levels, together with an extended way to characterize the amount of deposited energy along the transversal cross section of the detector described the LET spectra for both scenarios. With a wide LET(water) spectrum, from 0.5 to 8 keV·µm$^{-1}$, scattered protons represent the main particle class responsible for the out-of-field dose. Electrons and photons have a low LET(water) value, up to 1.5 keV·µm$^{-1}$. Not having a big impact on the biological effects on the healthy tissue behind Bragg peak, they must be considered to the total dose calculation. Regarding the ions and fast neutrons production, known as particles with high LET values with a maximum value at 6 keV·µm$^{-1}$, further experiments are needed to exploit the specific evidence of the metallic influence in the dose enhancement at the level of TPX3 detector.

Directional maps of scattered protons in both scenarios, with Ti implants and TE inserts, showed a detailed tracking distribution for two categories of protons: high and low energy respectively. In matter of high energy protons flux, there were not reported significant differences between those two irradiations. Instead, the scattered tracks of low energy protons produced by the TE implants showed up a more spread-out morphology, compared to the similar example, but when the metallic inserts were used.

Besides the study of the influence of implants in a proton treatment plan, distal to the SOBP, this study underlined the capabilities of the Timepix3 detectors to characterize the stray radiation.

Further work will focus on quantification and in deep evaluation of the morphological aspects for protons and neutrons. Moreover, the impact of the thermal neutrons on the out-of-field dose will be investigated. The specific tracks of thermal neutrons will overlap with those ones produced by protons. Using trained neural networks, the method will consist in an elaborate recognition process of particles based on spectral (deposited energy, LET) and directional distribution on the sensor surface. The dosimetric concepts for the out-of-field dose in proton-based treatment will further include artificial intelligence and machine learning algorithms to offer more information on the correlation between higher LET particles and the possibility of a secondary cancer induction.

**Acknowledgment**

This experiment has received financial support from project number 04-2-1132-2017/2022, founded from the JINR-Romania agreement. Work at ADVACAM was performed in the frame of Contract No. 4000130480/20/NL/GLC/hh from the 221 European Space Agency for the DPE project. We thank K. Shipulin and A. Agapov for their assistance and help during the experiment.